\documentclass[a4paper]{article}

\usepackage{INTERSPEECH2021}
\usepackage{multirow}
\usepackage{cite}
\usepackage{xcolor}
\usepackage{amsfonts}
\usepackage{subfig}

\title{An Improved Single Step Non-autoregressive Transformer for Automatic Speech Recognition}
\name{Ruchao Fan$^1$, Wei Chu$^2$, Peng Chang$^2$, Jing Xiao$^2$, Abeer Alwan$^1$}
\address{
  $^1$Dept. of Electrical and Computer Engineering, University of California, Los Angeles, USA\\
  $^2$PAII Inc., USA}
\email{\{fanruchao,alwan\}@g.ucla.edu, \{chuwei129,changpeng805, xiaojing661\}@pingan.com.cn}

\begin{document}

\maketitle
\begin{abstract}
  Non-autoregressive mechanisms can significantly decrease inference time for speech transformers, especially when the single step variant is applied. Previous work on CTC alignment-based single step non-autoregressive transformer (CASS-NAT) has shown a large real time factor (RTF) improvement over autoregressive transformers (AT). In this work, we propose several methods to improve the accuracy of the end-to-end CASS-NAT, followed by performance analyses. First, convolution augmented self-attention blocks are applied to both the encoder and decoder modules. Second, we propose to expand the trigger mask (acoustic boundary) for each token to increase the robustness of CTC alignments. In addition, iterated loss functions are used to enhance the gradient update of low-layer parameters. Without using an external language model, the WERs of the improved CASS-NAT, when using the three methods, are 3.1\%/7.2\% on Librispeech test clean/other sets and the CER is 5.4\% on the Aishell1 test set, achieving a 7\%$\sim$21\% relative WER/CER improvement. For the analyses, we plot attention weight distributions in the decoders to visualize the relationships between token-level acoustic embeddings. When the acoustic embeddings are visualized, we find that they have a similar behavior to word embeddings, which explains why the improved CASS-NAT performs similarly to AT.
\end{abstract}

\section{Introduction}

Transformers have been dominant in many sequence generation tasks, outperforming their recurrent neural network (RNN) counterparts in terms of both accuracy and speed for end-to-end systems\cite{vaswani2017attention, li2020comparison, karita2019comparative}. However, the autoregressive (left-to-right) generation order slows down inference speed significantly. To accelerate the inference, non-autoregressive transformers (NAT) were proposed for the parallel generation of the output sequence. The idea is widely adopted in neural machine translation (NMT)\cite{gu2018non, lee2020deterministic, saharia2020non}, automatic speech recognition (ASR)\cite{chen2020non, chan2020imputer, bai2020listen, tian2020spike, higuchi2020mask, fujita2020insertion, fan2021cass, chi2020align, song2021non, higuchi2021improved, fujita2020end, bai2021fast}, text-to-speech (TTS)\cite{peng2020non, miao2020flow} and speech translation \cite{inaguma2021orthros}. 

Current NAT models for ASR can be categorized into: (i) iterative NAT, and (ii) single step NAT, according to the number of iterations for sequence generation. Essentially, autoregressive models are also iterative-based since they use a left-to-right generation order and take N iterations to generate a sequence of length N. Hence, the idea of iterative NAT is to adopt a different generation order with less than N iterations to accelerate the inference. Chen et al. regarded the transformer decoder as a masked language model that first generates tokens with high confidence \cite{chen2020non}, while Higuchi et al. applied the same idea but based on the connectionist temporal classification (CTC) output \cite{higuchi2020mask,higuchi2021improved}. In addition, Fujita et al. used the idea of the insertion transformer from NMT to generate the output sequence with an arbitrary order \cite{fujita2020insertion}. Another recent effective method is using multiple decoders as refiners to do an iterative refinement based on CTC alignments \cite{chi2020align}. Theoretically, the iterative NAT has a limited improvement of inference speed since multiple iterations are still needed to obtain a competitive result. In contrast, single step NAT, which attempts to generate the output sequence with only one iteration, can have a better speed up for inference. The idea is to substitute the word embedding in autoregressive models with an acoustic representation for each output token, assuming that language semantics can also be captured by acoustic representations \cite{bai2020listen, tian2020spike, fan2021cass}.

Although various NAT methods were proposed for ASR, the WER performance still lags behind that of state-of-the-art autoregressive models. Therefore, based on our previous work\cite{fan2021cass}, we propose several methods to improve the accuracy of CTC alignment-based single step NAT (CASS-NAT) with little inference speed loss. First, convolution augmented self-attention blocks are applied to both the encoder and decoder modules, while other work only considered using them in the encoder\cite{gulati2020conformer}. The second method is to expand the trigger mask (acoustic boundary) for each token to increase the robustness of the CTC alignment. Third, considering the wide use of iterated loss functions to train deep transformers\cite{tjandra2020deja, wang2020transformer, lee2021intermediate}, we apply iterated loss to enhance the gradient update of low-layer parameters for both the encoder and decoder modules. When no external language model is used, large improvements are observed on both the Librispeech\cite{panayotov2015librispeech} and Aishell1\cite{bu2017aishell} datasets in terms of error rate, and the performance is close to the autoregressive baseline. Additionally, we analyse the self-attention distributions and token-level acoustic embeddings in the decoders. We find a similar behaviour between the token-level acoustic embedding and word embedding, which explains why CASS-NAT performs similarly to autoregressive models. 



The remainder of the paper is organized as follows. Section 2 briefly reviews the CASS-NAT and describes the proposed methods for improving the system. Section 3 describes the recognition experimental setup, followed by results and analyses in Section 4. Section 5 concludes the paper.

\section{System Overview}

\subsection{Basic CASS-NAT Model}
\label{ssec:review}

\begin{figure}[tp]
\centering
\centerline{\includegraphics[width=7.2cm,height=6.1cm]{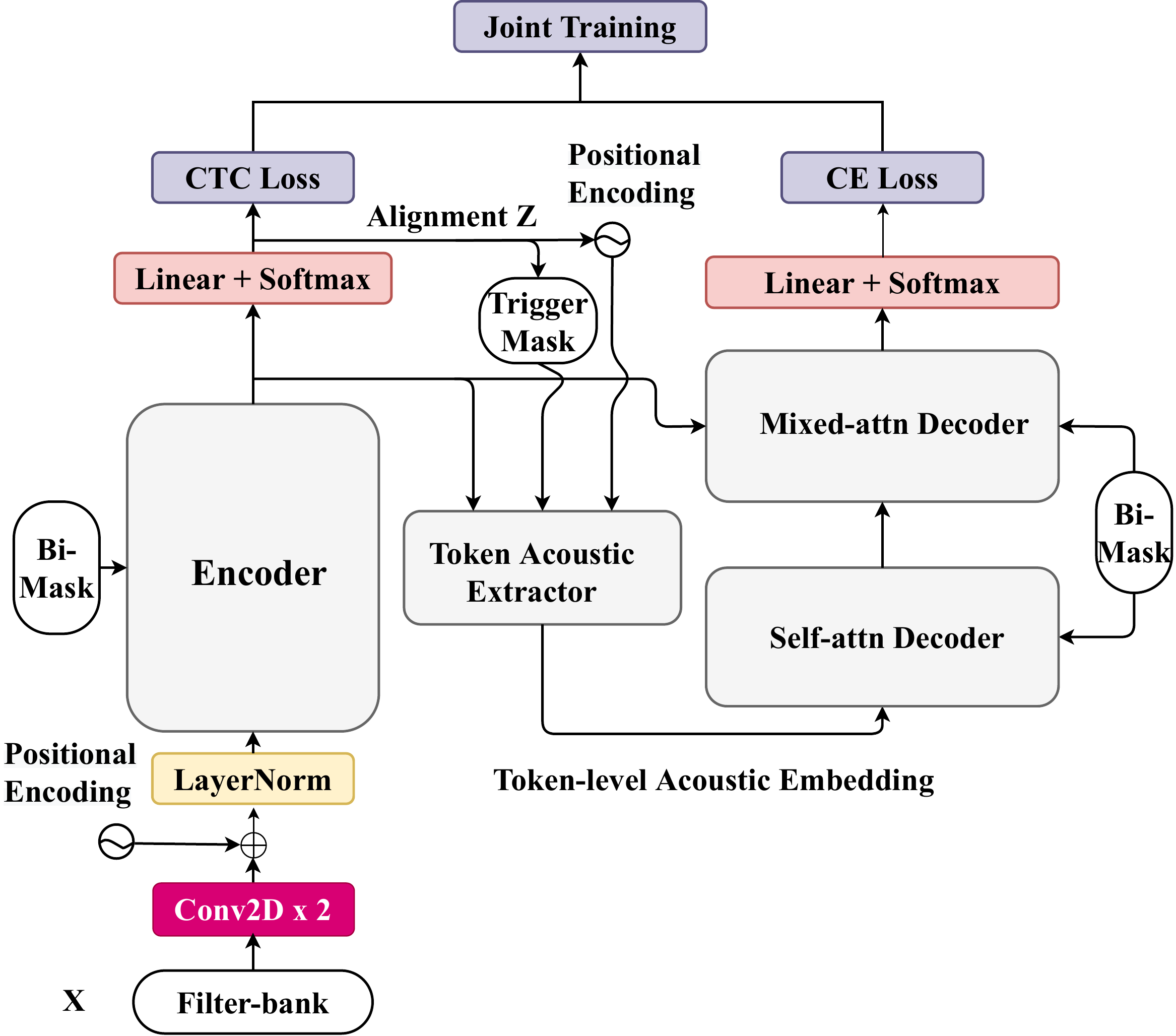}}
\caption{An overview of the CASS-NAT architecture\cite{fan2021cass}}
\label{fig:nast}
\end{figure}

The CTC alignment-based single step non-autoregressive transformer (CASS-NAT) that we proposed in \cite{fan2021cass} is modified based on the CTC/Attention hybrid architecture\cite{watanabe2017hybrid} to be non-autoregressive. Fig.\ref{fig:nast} shows four major modules in the CASS-NAT: encoder, token acoustic extractor (TAE), self-attention decoder (SAD) and mixed-attention decoder (MAD).

Suppose the input sequence is $X=\{x_1,...,x_t,...,x_T\}$ and ground truth is $Y=\{y_1,...y_u,...,y_U\}$, the encoder extracts high level acoustic representations $H$ from speech features $X$, followed by a CTC loss function. The role of CTC is to obtain an alignment over CTC output space to offer auxiliary information for the token acoustic extractor. Specifically, given an alignment $Z=\{z_1,...,z_t,...,z_T\}$, we can estimate an acoustic segment for each token $u$ as $\{z_{t_{u-1}+1},...z_{t_u}\}$, as well as the number of tokens in $Z$. The segment is equivalent to the trigger mask in Fig. \ref{fig:nast}, which is used for self-attention computation. The information obtained from the CTC alignment is then used to extract token-level acoustic embeddings that replace word embeddings existing in autoregressive transformers (AT). The self-attention and mixed-attention decoders are finally used to model the relationship between tokens, where the MAD has access to the acoustic representations $H$, while the SAD does not.

The framework is trained by jointly optimizing a CTC loss function on the encoder side ($L_{CTC}$) and a cross entropy (CE) loss function on the decoder side ($L_{dec}$) so that the final loss function ($L_\text{joint}$) is defined as: 
\begin{equation}
    L_{\text{joint}} = \lambda \cdot \log \sum_{Z\in q}{\prod_{i=1}^T{P(z_i|X)} + \log\prod_{u=1}^U{P(y_u|z^{*}_{t_{u-1}+1:t_{u}}, X)}}
\label{eq:cassloss}
\end{equation}
where P is the probability, $\lambda$ is the task ratio of the CTC loss, and q is a set of alignments that can be mapped to the ground truth $Y$. $Z^*$ is the most probable alignment, obtained by forced alignment over the CTC output space. Note that the second term is a maximum approximation for the loss on the decoder side.

\subsection{Convolution Augmented Self-attention Block}

Self-attention is the most basic layer in speech transformers, and is formulated as:
\begin{equation}
\label{eq:attention}
    Attn(Q, K, V, M) = Softmax(\frac{QK^T}{\sqrt{d_k}})\otimes M \cdot V
\end{equation}
where $Q \in R^{n_q\times d_q}$, $K \in R^{n_k\times d_k}$, $V \in R^{n_v\times d_v}$, and $M \in R^{n_q \times n_k}$ are the queries, keys, values and mask matrix, respectively. From the equation, it can be seen that self-attention considers global information across the sequence, but ignores local details. To alleviate this problem, convolution augmented self-attention blocks are proposed to emphasise the modelling of local dependencies of the input sequence in the encoder \cite{gulati2020conformer, yang-etal-2019-convolutional}. Different from previous work, we apply the convolution augmented self-attention blocks in the mixed-attention decoder as well as the encoder. Specifically, the feed-forward layer is decomposed into two sub-layers to be placed at the beginning and the end of the block. A convolution layer similar to that in \cite{gulati2020conformer} is inserted after the self-attention layer except that we empirically use layer normalization instead of batch normalization. The final computation in the $i^{th}$ MAD can be formulated as:

\begin{align}
\label{eq:mad_block}  
    \hat{s_i} &= s_i + \frac{1}{2}\text{FFN}(s_i) \\
    s_i^{'} &= \hat{s_i} + \text{LN}(\text{Attn}(\hat{s_i}, \hat{s_i}, \hat{s_i}, \text{BiMask})) \\
    s_i^{''} &=  s_i^{'} + \text{Conv}(s_i^{'}) \\
    s_i^{'''} &= s_i^{''} + \text{LN}(\text{Attn}(s_i^{''}, H, H, \text{BiMask})) \\
    o_i &= \text{LN}(s_i^{'''} + \frac{1}{2}\text{FFN}(s_i^{'''}))
\end{align}
where LN indicates layer normalization and FFN is the feed-forward layer. BiMask stands for a bidirectional mask.

In addition, different from the usage of relative positional encoding in \cite{dai2019transformer,gulati2020conformer}, we consider a maximum length of relative position k as in \cite{zhou2019improving}. Therefore, $2k+1$ position embedding are learned to represent the relative position between $[-k, k]$.

\subsection{Trigger Mask Expansion}
The quality of token-level acoustic embedding relies on the accuracy of the trigger mask, which is mapped from the CTC alignment. Although the CTC loss function is used to optimize the alignment, there are still errors when doing forced alignment over CTC output space, leading to an inaccurate trigger mask. In order to compensate the inaccuracy of token-level acoustic embedding extraction, we expand the trigger mask to include contextual frames for each token. For example, suppose the contextual frame size is one, the acoustic boundary of token U becomes $\{z_{t_{u-1}},...,z_{t_u+1}\}$. The trigger mask will then be expanded by one in the subsequent acoustic embedding extraction.

\subsection{Iterated Loss}
Deep transformers always suffer from gradient vanishing, especially for parameters that are distant from the output layers. Iterated loss are proposed to add additional loss functions after each layer to boost the gradient update\cite{szegedy2015going,tjandra2020deja}. Recent work proposed iterated CTC loss to improve the performance of CTC\cite{lee2021intermediate}. In \cite{wang2020transformer}, iterated CE loss is used for training deep transformer-based acoustic models. Since CTC and CE loss functions are jointly optimized in the CASS-NAT framework, we integrate iterated CTC and CE loss functions into Eq.\ref{eq:cassloss} so that the parameters in different layers can be updated at the same scale. We find this strategy to be more effective for CASS-NAT than AT models. Let $L_{dec}=\log\prod_{u=1}^U{P(y_u|z^{*}_{t_{u-1}+1:t_{u}}, X)}$ and $L_{CTC}=\log \sum_{Z\in q}{\prod_{i=1}^T{P(z_i|X)}}$, the objective function is re-written as:
\begin{equation}
\begin{aligned}
    L_{\text{joint}} &= \lambda_{CE}L_{dec}^{final} + (1-\lambda_{CE})L_{dec}^{middle} \\
    &+\lambda_{CTC}L_{CTC}^{final} + (1-\lambda_{CTC})L_{CTC}^{middle}
\end{aligned}
\end{equation}
where $\lambda_{CE}$ and $\lambda_{CTC}$ are task ratios. Middle and final indicate the layer position of the inserted loss functions.

\section{Experimental Setup}
\begin{table}[t]
\caption{WERs of the proposed methods for improving CASS-NAT on Librispeech. No external language model is used. SpecAug is used in all configurations. WERR is the incremental relative WER improvement on the test-other data. }
\footnotesize
\centering
\begin{tabular}{l cccc c}
\hline
\multirow{2}{*}{Model w/o LM} & dev- &  dev- &  test- &  test- & \multirow{2}{*}{\textit{WERR}} \\
~ & clean &  other &  clean &  other & ~ \\
\hline\hline
Conformer AT & 2.7 & 7.2 & 3.0 & 7.0 & ~\\
\hline
CASS-NAT & 3.7 & 9.2 & 3.8 & 9.1 & - \\
+ Conv-aug Enc. & 3.1 & 7.9 & 3.3 & 7.9 & 13.2\% \\
+ Conv-aug Dec. & 3.0 & 7.8 & 3.1 & 7.6 & 3.8\% \\
+ Tri. Mask Exp. & 3.0 & 7.6 & 3.1 & 7.5 & 1.3\% \\
+ Iterated CTC & 2.8 & 7.3 & 3.1 & 7.3 & 2.7\% \\
+ Iterated CE & 2.8 & 7.3 & 3.1 & 7.2 & 1.4\% \\
\hline
\end{tabular}
\label{tab:improvements_libri}
\end{table}

The experiments were conducted on three datasets: the 960-hour LibriSpeech English corpus\cite{panayotov2015librispeech}, the 178-hour Aishell1 Mandarin corpus\cite{bu2017aishell} and the 50-hour OGI Kids corpus (English)\cite{shobaki2000ogi}. All experiments used 80-dim Mel-filter bank features, computed every 10ms with a 25ms Hamming window. Features of every 3 consecutive frames were concatenated to form a 240-dim feature vector as the input. The sets of output labels consist of 5k word pieces obtained by the SentencePiece method\cite{kudo2018sentencepiece} for Librispeech and OGI. 4230 Chinese characters were obtained from the training set for the Aishell1 dataset.

A CTC/Attention AT baseline was first trained with the architecture ($N_e=10$, $N_d=5$, $d_{ff}=2048$, $nh=8$, $d_{att}=512$) for Librispeech, ($N_e=12$, $N_d=6$, $d_{ff}=2048$, $nh=4$, $d_{att}=256$) for the Aishell1 and ($N_e=8$, $N_d=4$, $d_{ff}=2048$, $nh=4$, $d_{att}=256$) for the OGI. When training the CASS-NAT, the encoder was initialized with the AT encoder for faster convergence as in \cite{fan2019online}. The decoder in the AT baseline was replaced by 1-block token-level acoustic extractor, 3-block self-attention decoder and 4-block mixed-attention decoder. The maximum length of relative position k was set to 20 in the encoder and 8 in the decoder for the English tasks and 4 in the decoder for the Aishell1 data. The contextual frame of the trigger mask expansion was 1. The iterated loss functions were inserted in the middle layer of the encoder and mixed-attention decoder with $\lambda_{CE}$ of 0.9 and $\lambda_{CTC}$ of 0.5. The inserted projection layers were not used during inference. These settings were empirically chosen based on many experiments.
 
Each of the two convolution layers in Fig.\ref{fig:nast} has 64 filters with a kernel size of $3\time 3$ and a stride of 2, leading to a 4x frame-rate reduction. The same learning schedule in \cite{fan2021cass} was adopted. Layer normalization, dropout with rate of 0.1 and label smoothing with a penalty of 0.1 were all applied as the common strategies for training a transformer. We also applied SpecAug\cite{park2019specaugment} for fair comparisons with results in previous literature. We additionally applied speed-perturbation for the Aishell1 dataset for fair comparisons with previously published results. We used development sets for early stopping and model averaging for final evaluation. Most of the experiments ended within 90 epochs.
During AT decoding, the beam size is set to 20 for Librispeech, and it is set to 10 for the Aishel1 and OGI. The evaluation of the real time factor (RTF) was conducted using an NVIDIA Tesla V100 GPU with batch size of one. 

A transformer-based language model was trained with the provided text in Librispeech for shallow fusion during AT baseline decoding. The language model was also used for ranking alignments and beam search during CASS-NAT decoding.
\section{Results and Analyses}

\subsection{Results on Librispeech and Aishell1}
\label{ssec:results}
Experiments are first conducted on the Librispeech corpus by incrementally adding the proposed methods based on the original CASS-NAT\cite{fan2021cass} as shown in Table \ref{tab:improvements_libri}. First, when convolution augmented self-attention blocks are applied to the encoder, the WER on the test-other set has a 13.2\% relative improvement compared to the CASS-NAT baseline. When the convolution augmented blocks are also applied to the decoder, the WER drops further by a 3.8\% relative improvement. An 1.3\% relative WER improvement is observed for the trigger mask expansion method. When using the iterated loss function to both the encoder and decoder, we obtain an incremental 2.7\% and 1.4\% WER improvements, respectively. The final result has less than a 3\% increase in relative WER compared to the AT baseline, which used a conformer structure. In the next sub-section, we will explain why the improved CASS-NAT can achieve a performance that is close to its autoregressive counterpart.

\begin{table}[t]
\caption{A comparison of error rates and RTFs with previously published results. RTFs of previous work are missing because the authors did not report them, or the machines used to test RTFs are different. $\dagger$: use SpecAug. $\star$: use speed perturbation.}
\footnotesize
\centering
\begin{tabular}{l  c  c  c  c}
\hline
\multirow{2}{*}{Librispeech (WER)} & \multirow{2}{*}{LM} & test & test & RTF \\
~ & ~ & clean & other & test-other \\
\hline\hline
\multicolumn{5}{l}{Previous work (NAT)} \\
A-FMLM \cite{chen2020non} $\dagger$ & w/o & 6.6 & 12.2 & - \\
Imputer\cite{chan2020imputer} & w/o & 4.0 & 11.1 & - \\
Align-refine\cite{chi2020align} $\dagger$  & w/o & 3.6 & 9.0 & - \\
CASS-NAT\cite{fan2021cass} $\dagger$ & w/o & 3.8 & 9.1 & 0.010 \\
\hline
\multirow{2}{*}{$\text{Conformer AT}^\dagger$}  & w/o & 3.0 & 7.0 & 0.499 \\
~ & w/ & 2.3 & 5.2 & 0.568 \\
\multirow{2}{*}{Improved CASS-NAT} $\dagger$ & w/o & \textbf{3.1} & \textbf{7.2} & 0.014 \\
~ & w/ & 2.8 & 6.5 & 0.188 \\
\hline \hline
Aishell1 (CER) & LM & dev & test & RTF test \\
\hline \hline
\multicolumn{5}{l}{Previous work (NAT)} \\
ST-NAT \cite{tian2020spike} $\dagger$ & w/o & 6.9 & 7.7 & - \\
A-FMLM \cite{chen2020non} $\star$ & w/o & 6.2 & 6.7 & - \\
Insertion-NAT \cite{fujita2020insertion} $\dagger$ & w/o & 6.1 & 6.7 & - \\
Enhanced-NAT\cite{song2021non} $\dagger \star$ & w/o & 5.3 & 5.9 & - \\
BERT-LASO \cite{bai2021fast} $\dagger$ & w/o & 5.2 & 5.8 & - \\
CASS-NAT\cite{fan2021cass} $\dagger\star$ & w/o & 5.3 & 5.8 & 0.011 \\
\hline
Conformer AT $\dagger\star$ & w/o & 4.8 & 5.2 & 0.200 \\
Improved CASS-NAT $\dagger\star$ & w/o & \textbf{4.9} & \textbf{5.4} & 0.023 \\
\hline
\end{tabular}
\label{tab:results2}
\end{table}

The proposed three methods are also used to train an improved CASS-NAT on the Aishell dataset. The final WER and real time factor (RTF) comparisons with previously published results are shown in Table \ref{tab:results2}, including both the Librispeech and Aishell1 data. Using the Librispeech dataset, the WER on test-other for the proposed methods has a 21\% relative improvement over the original CASS-NAT with little RTF degradation (from 0.010 to 0.014). This RTF still has a 36x speed up compared to the AT baseline when no external LM is used. When using an external LM during inference, CASS-NAT does not benefit from LM as much as the AT baseline does, which was also reported in \cite{fan2021cass}. Using the Aishell1 dataset, the RTF speed up is not as large as in Lirbispeech. The reason may be that the speed up of NAT benefits longer utterances; however, there are more short utterances in Aishell1. In addition, we use AT baseline for ranking alignments, which may increase the computational cost. Although we do not carefully tune the hyper-parameters on Aishell1, the character error rate (CER) on the test set still improves from 5.8\% to 5.4\%, which is close to the AT baseline.

\subsection{Attention and Embedding Visualizations}
\label{ssec:attplot}

\begin{figure}[t!]
    \centering
    \subfloat[SAD head 5]{{\includegraphics[width=2cm,height=1.5cm]{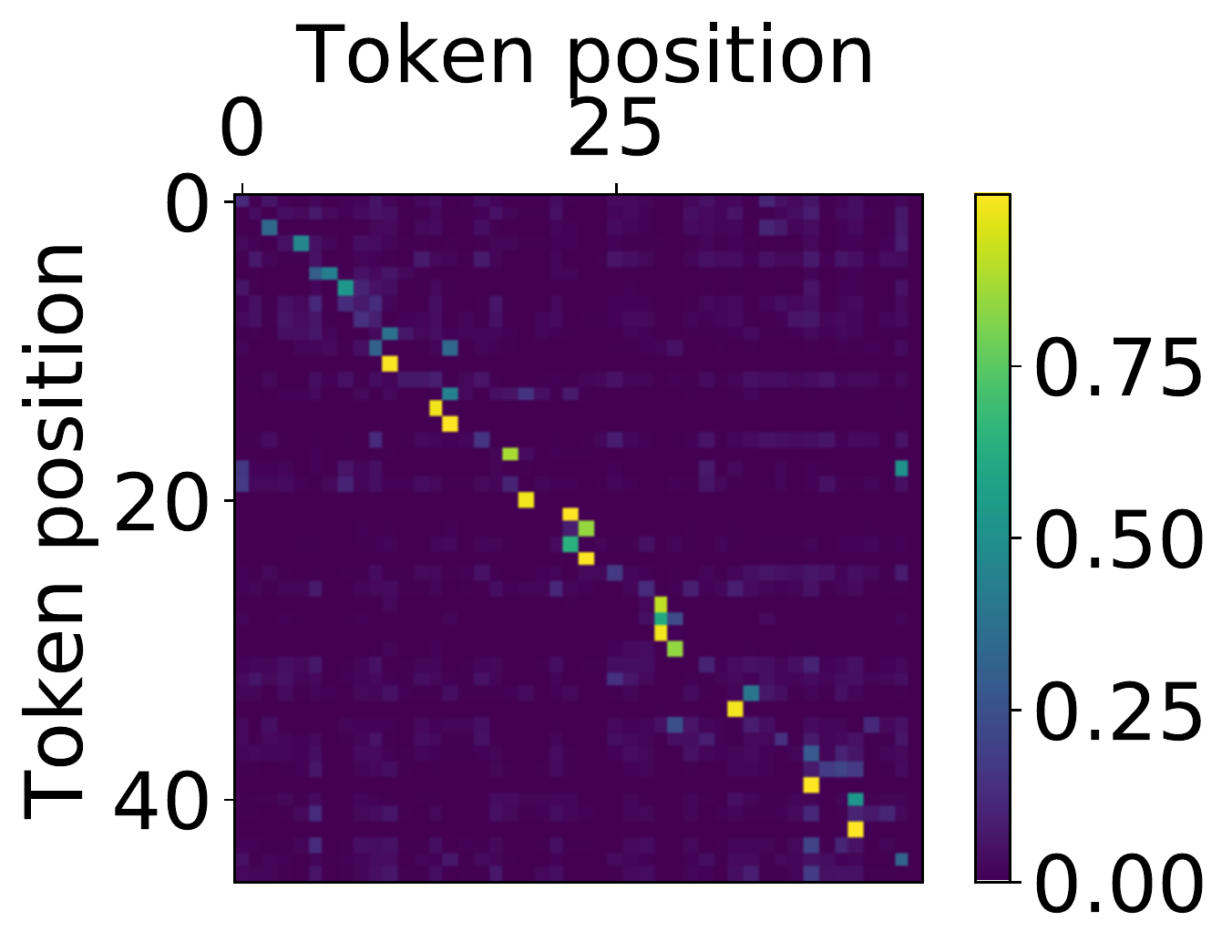} }} %
    \subfloat[SAD head 6]{{\includegraphics[width=2cm,height=1.5cm]{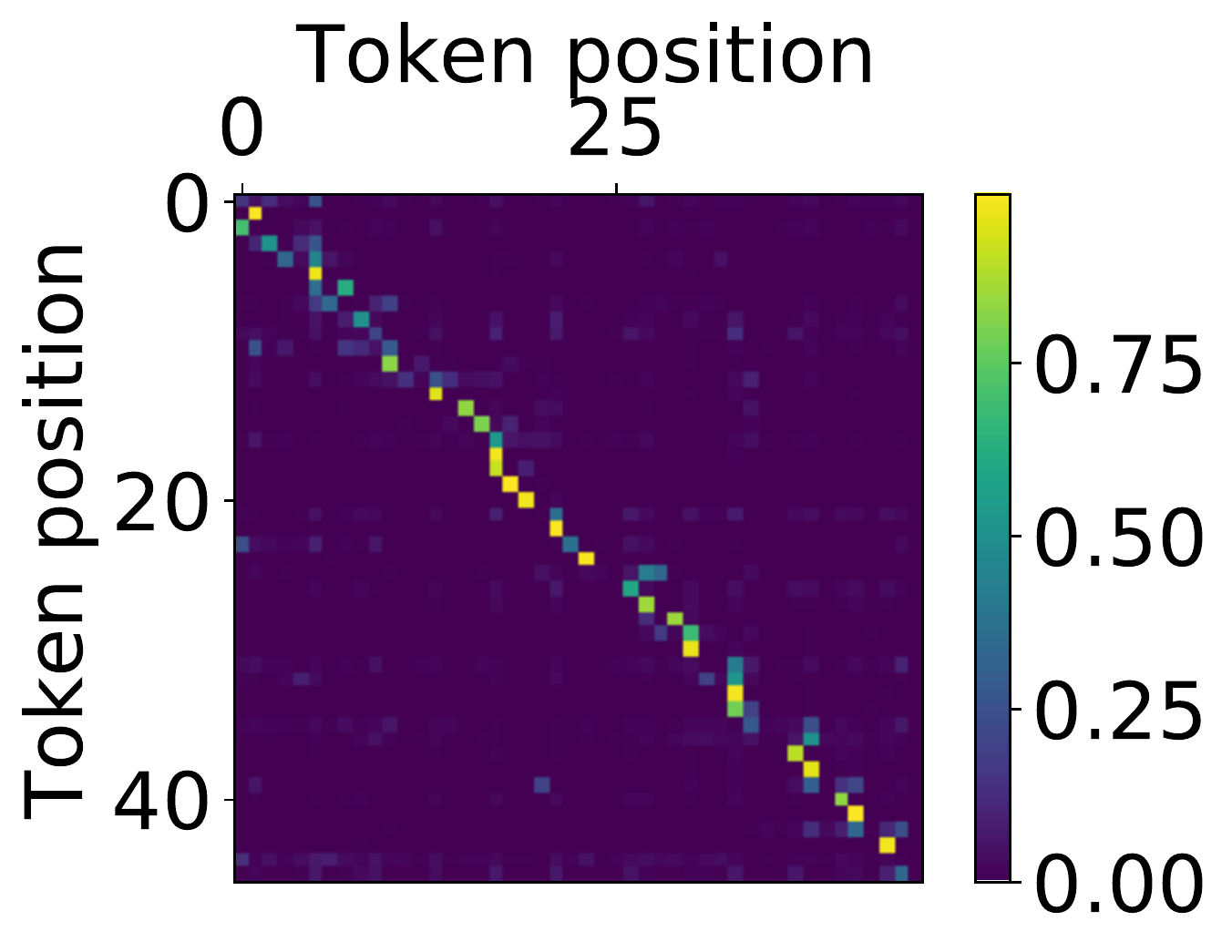} }} %
    \subfloat[SAD head 7]{{\includegraphics[width=2cm,height=1.5cm]{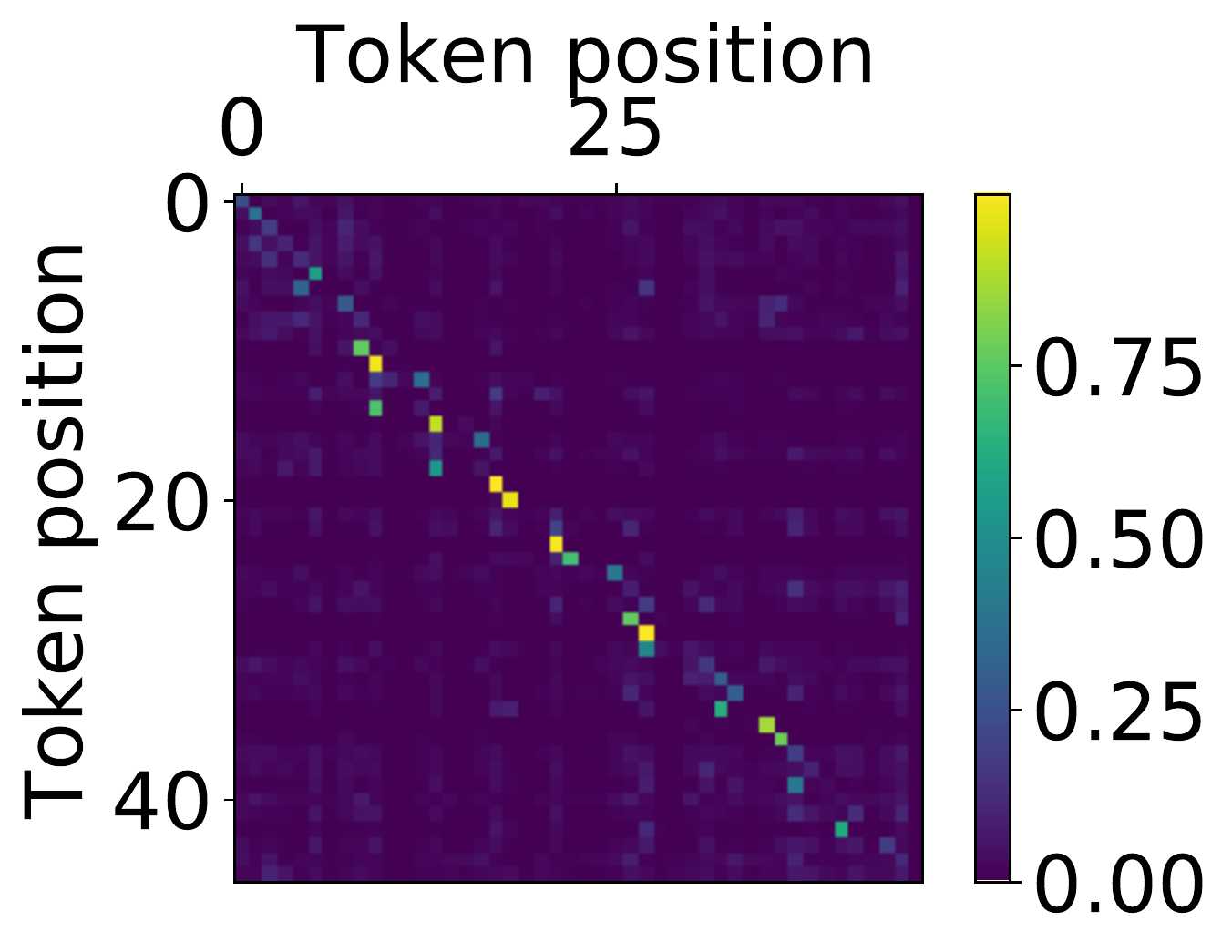} }} %
    \subfloat[SAD head 8]{{\includegraphics[width=2cm,height=1.5cm]{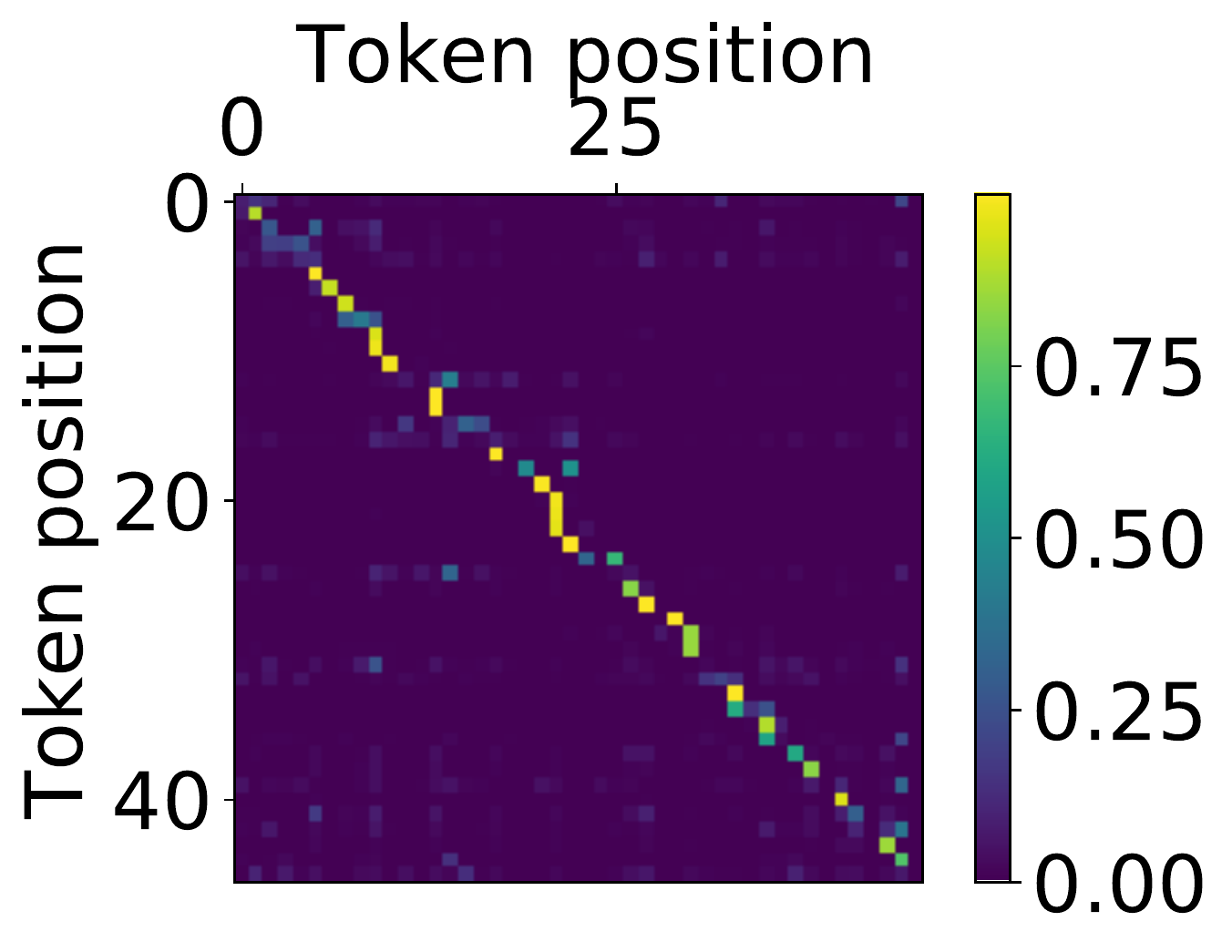} }} %
    \qquad
    \subfloat[MAD head 5]{{\includegraphics[width=2cm,height=1.5cm]{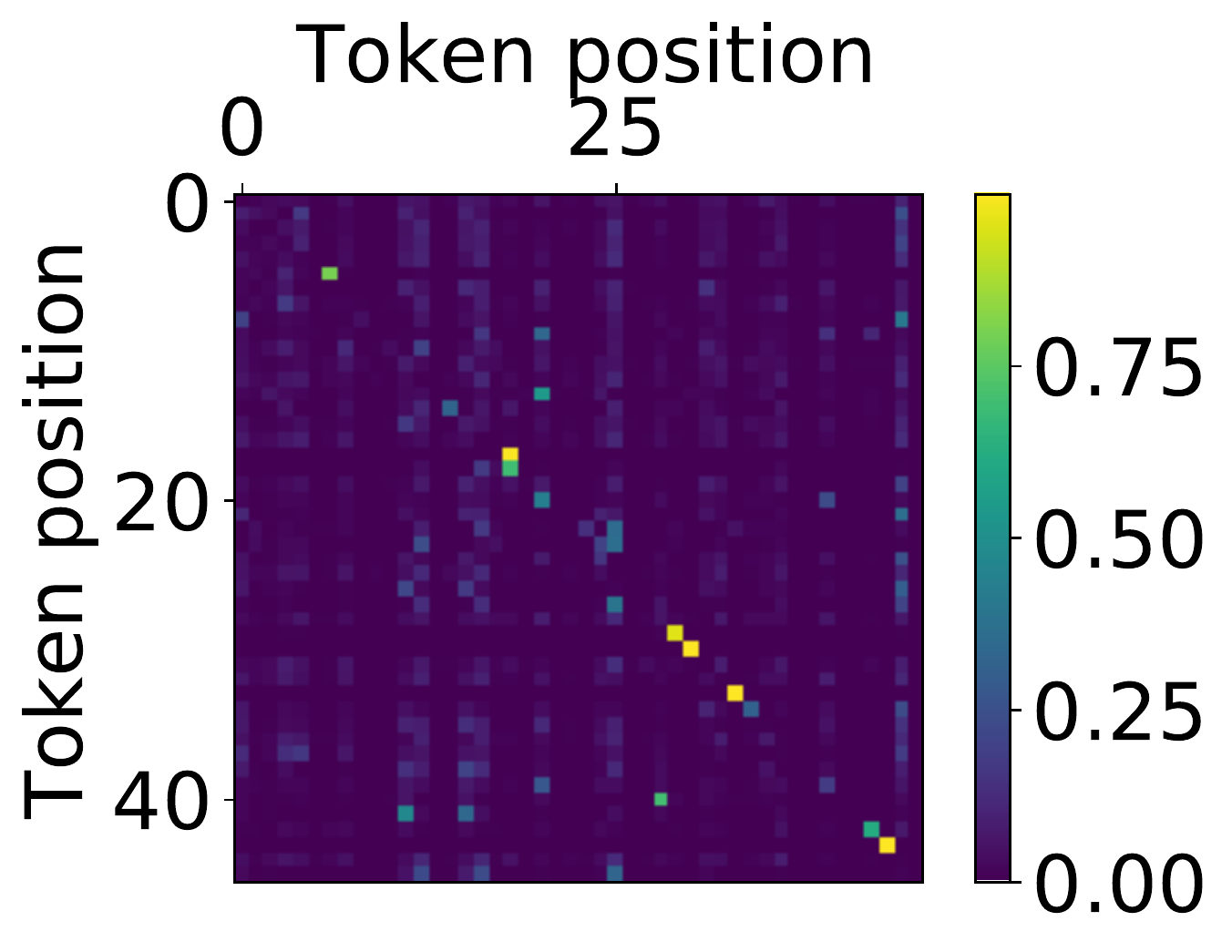}}} %
    \subfloat[MAD head 6]{{\includegraphics[width=2cm,height=1.5cm]{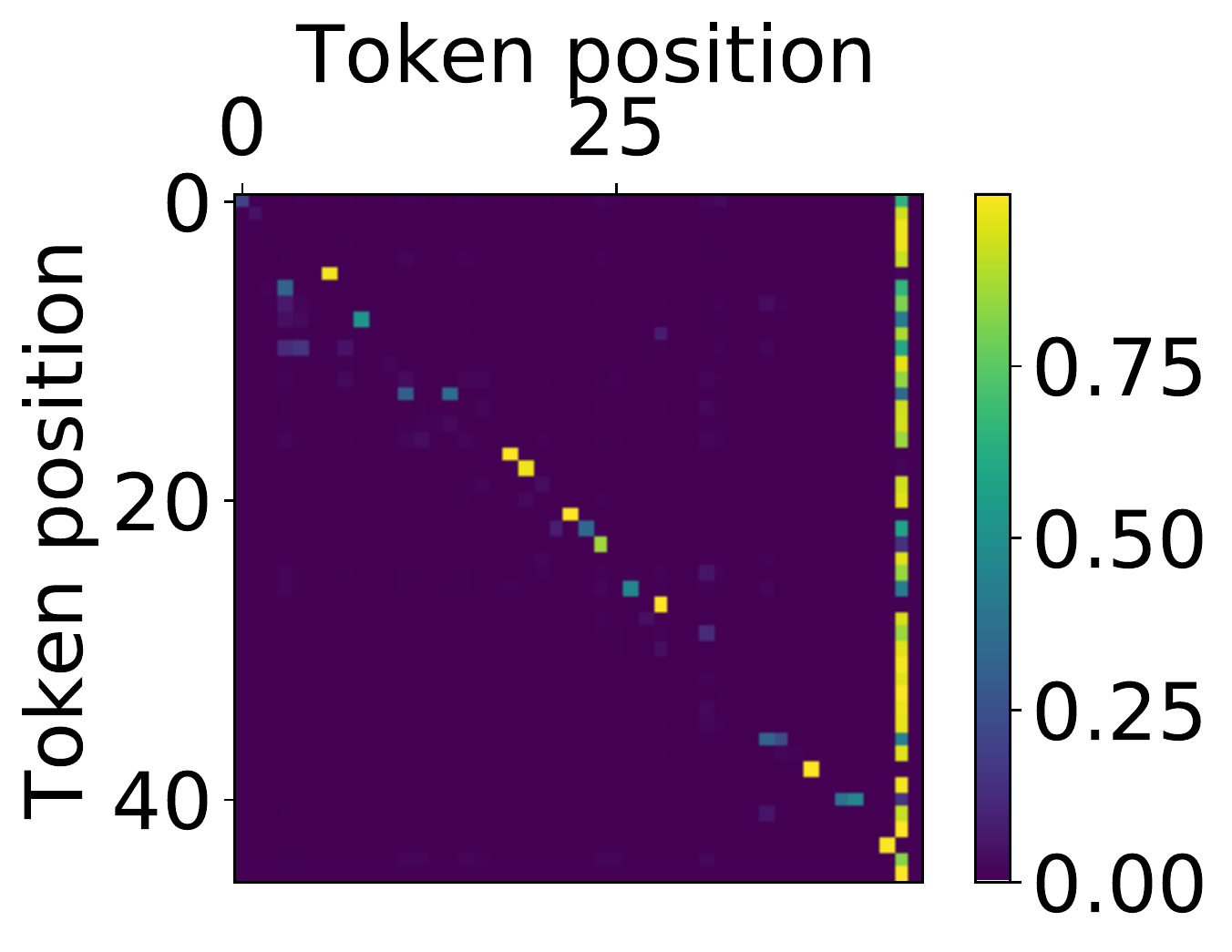} }} %
    \subfloat[MAD head 7]{{\includegraphics[width=2cm,height=1.5cm]{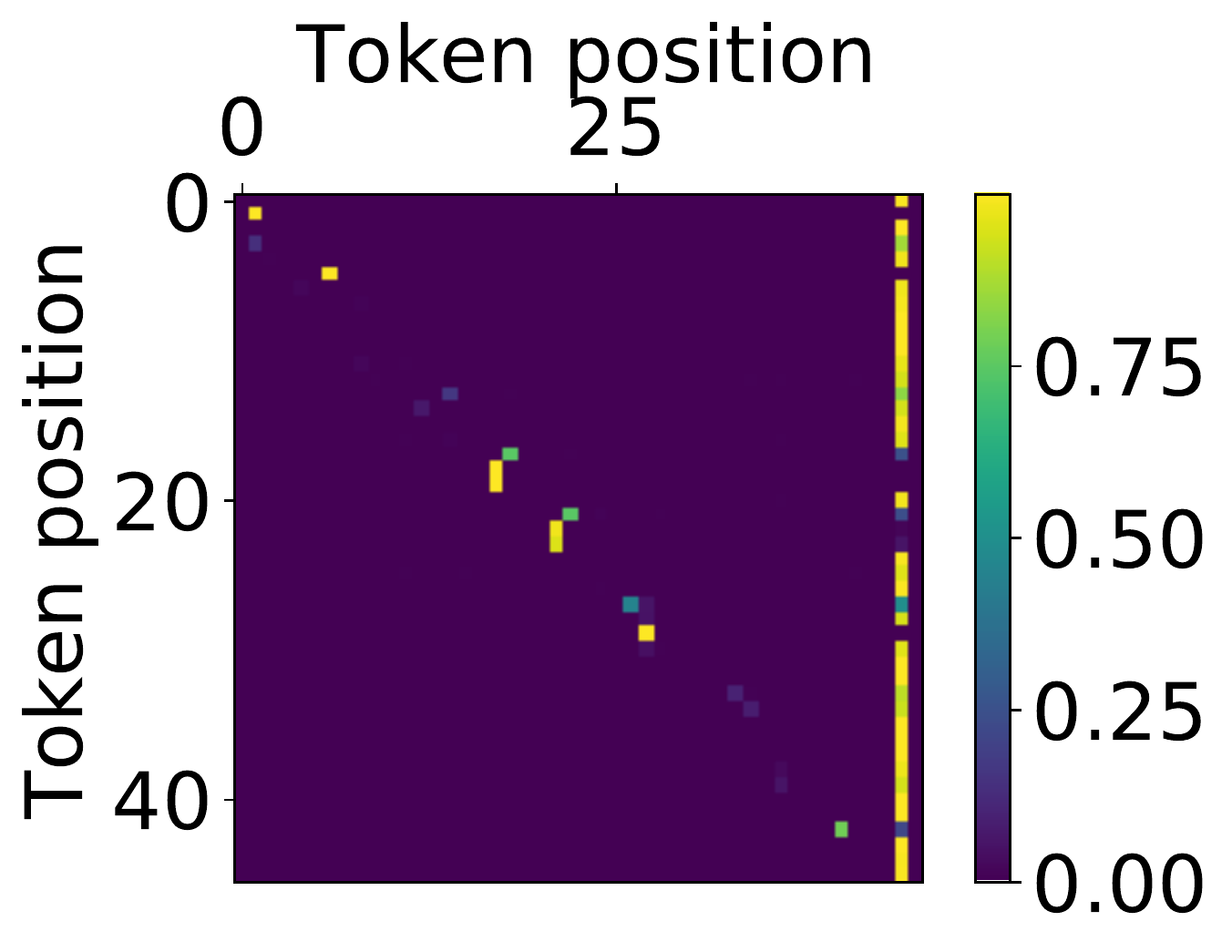} }}  %
    \subfloat[MAD head 8]{{\includegraphics[width=2cm,height=1.5cm]{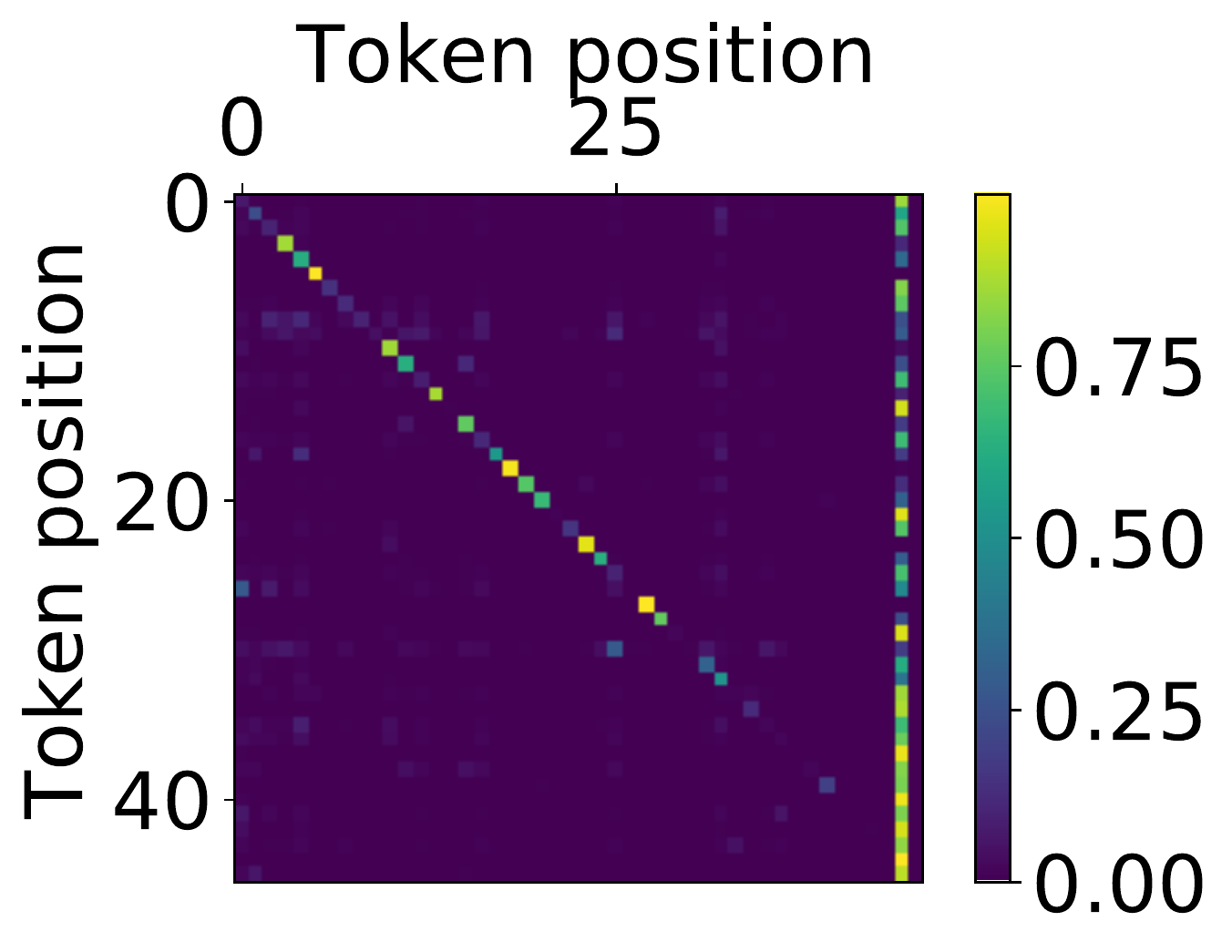} }} %
    \caption{Attention weight distributions of the last four heads in multi-head self-attention of the last block in the self-attention decoder (SAD) and mixed-attention decoder (MAD) for the first utterance in the Lirispeech train-clean-100 subset. The matrices $Q,K,V$ in Eq.\ref{eq:attention} are split into ${nh}$ sub-spaces and then they are used for self-attention computation separately. Each subspace is referred to as a head. $nh$ is set to 8 for this dataset.} %
    \label{fig:attweight}%
\end{figure}

\begin{figure}[t!]
    \centering
    \subfloat{{\includegraphics[width=7cm,height=3cm]{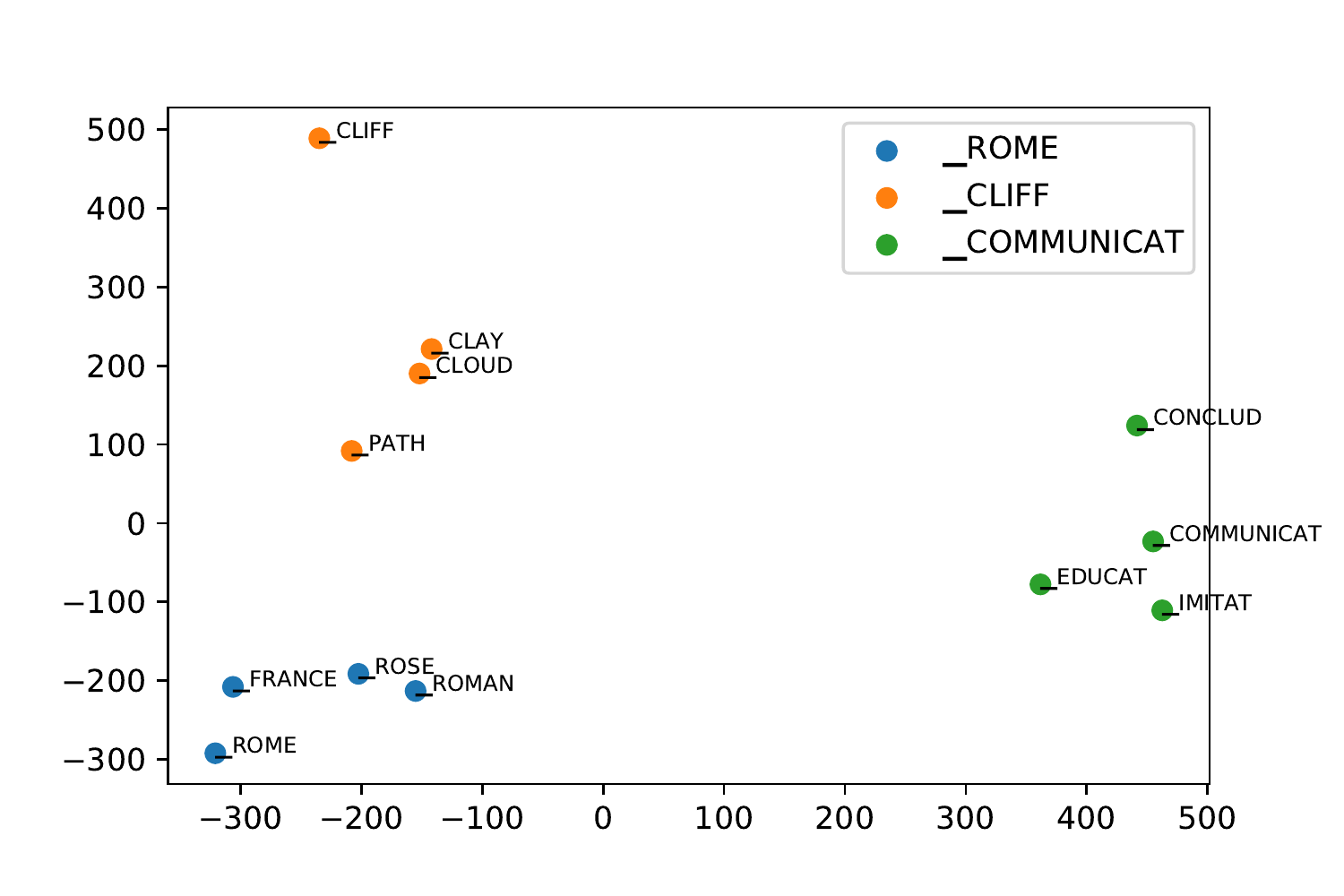} }}%
    \qquad
    \subfloat{{\includegraphics[width=7cm,height=3cm]{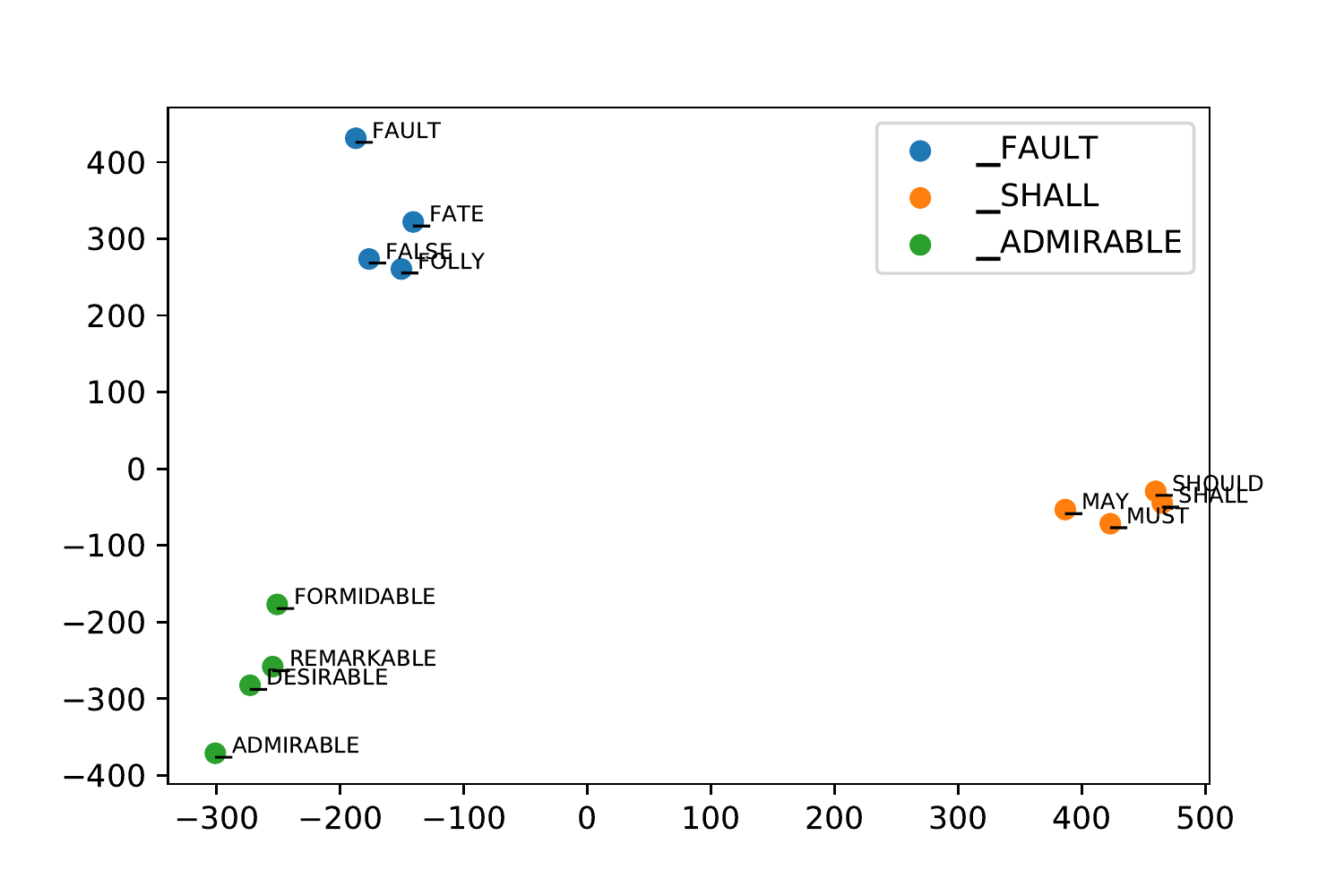}}}%
    \caption{Visualization of token-level acoustic embedding for six word pieces using the first two dimensions after PCA.}%
    \label{fig:acembed}
\end{figure}

To analyse the modelling of token-level acoustic embeddings, we choose the first utterance in the Librispeech train-clean-100 subset and plot the self-attention weights in the last block of the self-attention and mixed-attention decoders. The weights of the last 4 heads are shown in Fig.\ref{fig:attweight}. We can see from the figure that most of the heads learn a monotonic alignment between the token-level acoustic embeddings, indicating that each token relies more on adjacent tokens, which is similar to the idea of word embedding using continuous bag of words (CBOW) and skip-gram \cite{mikolov2013efficient}. The monotonic alignment also shows the usefulness of the relative positional encoding because distant tokens with close semantic similarity have low attention weights. 

To further investigate how the token-level acoustic embedding behaves, we extracted such embeddings from the first 3000 utterances in the train-clean-100 Librspeech subset. Each acoustic embedding has its own word piece ground truth. For each word piece, the embeddings are averaged to represent the final acoustic embedding. Using the same idea of visualizing word embedding, we randomly choose three word pieces and find their four closest embeddings using the cosine similarity distance. The 12 embeddings are reduced to a 2-dimensional space using principal component analyses (PCA) and are then plotted. Two examples are shown in Fig.\ref{fig:acembed}. The figure shows that the token-level acoustic embedding can learn not only the acoustic similarity, but also the word-piece level semantic similarity. For example, for the word piece \textbf{\_ROME}, the closest tokens contain \textbf{\_FRANCE} representing cities and countries, and \textbf{\_ROSE} which has a similar pronunciation. The same situation can be seen in the lower part of the Fig.\ref{fig:acembed} where each cluster has the same part-of-speech words, such as \textbf{\_MAY} versus \textbf{\_SHOULD} and \textbf{\_REMARKABLE} versus \textbf{\_FORMIDABLE}. This behaviour of token-level acoustic embedding is very similar to word embedding, indicating that they can also be used to capture semantics. This may explain why the improved CASS-NAT has a similar performance to the AT baseline.

\subsection{NAT for Child Speech}
\label{ssec:nat_kids}

\begin{table}[th]
\caption{WER for the development and test sets and RTF for the test set using the scripted part of the OGI corpus. Both experiments used SpecAug.}
\centering
\begin{tabular}{l cc c}
\hline
~ & dev &  test & RTF on test\\
\hline\hline
Conformer AT & 1.8 & 2.5 & 0.081\\
\hline
Improved CASS-NAT & 2.2 & 2.6 & 0.018 \\
\hline
\end{tabular}
\label{tab:ogi}
\end{table}

In this section, we conduct experiments on the scripted part of the OGI kids corpus with the data partition in \cite{fan2021bi} except that 10\% of the test set is chosen to be a development set for early stopping. The results are shown in Table \ref{tab:ogi}. Since the OGI dataset has a fixed language pattern, the distributions of training and test sets are close. Although we trained the model within 15 epochs, the WER remains very competitive. Nevertheless, the conclusion of our improvements to CASS-NAT still holds and it has an impressive speed up than autoregressive models, which may be suitable for child ASR-based educational applications. 

\section{Summary and Conclusion}

This paper presented three methods to improve the WER performance of CTC alignment-based single step NAT (CASS-NAT), followed by performance analyses. First, convolution augmented self-attention blocks were applied to the encoder and decoder modules. Second, the trigger mask was expanded for each token to compensate for the inaccuracy of CTC alignments. Third, iterated loss functions were used to enhance the gradient update of low-layer parameters. When using the three methods without external language models, we achieved a 7\%$\sim$21\% relative WER/CER improvement over the original CASS-NAT on the Librispeech and Aishell1 dataset with little RTF degradation. The WER performance was worse within 5\%, in relative terms, than the autoregressive baseline, but maintained a 10$\sim$40x speed up. Moreover, attention weights and embedding visualization showed that the token-level acoustic embedding had similar behaviors with word embedding, explaining why the CASS-NAT performed similarly to AT.

\section{Acknowledgements}
This work was supported in part by the NSF.

\bibliographystyle{IEEEtran}

\bibliography{mybib}
\end{document}